\documentclass[aps,twocolumn,preprint,showpacs,
amsmath,amssymb]{revtex4}

\usepackage{etex}

\usepackage{amsthm,amscd,amsbsy,array}
\usepackage{bm}
\usepackage{soul} 

\usepackage{graphics,graphicx,xcolor}

\usepackage[colorlinks=true, pdfstartview=FitV, linkcolor=blue, citecolor=blue, urlcolor=blue]{hyperref} 

\numberwithin{equation}{section}

\let\ssection=\section
\renewcommand{\section}{\setcounter{equation}{0}\ssection}

\newcommand{\bb}{{\bf b}}

\newcommand{\bbeta}{\boldsymbol{\beta}}

\newcommand{\NU}{\mathrm{NU}}

\newcommand{\bbC}{\mathbb{C}}

\newcommand{\BMS}{{\mathrm{BMS}}}

\newcommand{\CCarr}{{\mathrm{CCarr}}}
\newcommand{\Carr}{{\mathrm{Carr}}}
\newcommand{\carr}{{\mathfrak{carr}}}

\newcommand{\confcarr}{{\mathfrak{ccarr}}}
\newcommand{\hGamma}{{\widehat{\Gamma}}}

\newcommand{\Diff}{{\mathrm{Diff}}}

\newcommand{\rg}{\mathrm{g}}

\newcommand{\hrg}{\widehat{\rg}}

\newcommand{\bgamma}{\boldsymbol{\gamma}}

\newcommand{\cI}{{\mathcal{I}}}

\newcommand{\bkappa}{\boldsymbol{\kappa}}

\newcommand{\rO}{{\mathrm{O}}}

\newcommand{\bomega}{{\boldsymbol{\omega}}}

\newcommand{\PSL}{\mathrm{PSL}}

\newcommand{\hg}{\hat{g}}

\newcommand{\bx}{{\bm{x}}}

\newcommand{\bbR}{\mathbb{R}}

\newcommand{\SL}{\mathrm{SL}}

\newcommand{\SO}{\mathrm{SO}}

\newcommand{\soo}{\mathfrak{so}}

\newcommand{\hX}{Y
}
\newcommand{\Conf}{{\mathrm{Conf}}}
\newcommand{\conf}{{\mathrm{conf}}}
\newcommand{\cT}{\mathcal{T}}

\newcommand{\bz}{{\bf z}}

\newcommand{\bbZ}{\mathbb{Z}}

\def\smallover#1/#2{\hbox{$\textstyle\frac{#1}{#2}$}} %

\def\beq{\begin{equation}}
\def\eeq{\end{equation}}
\def\beqa{\begin{eqnarray}}
\def\eeqa{\end{eqnarray}}

\def\barray{\left(\begin{array}}
\def\earray{\end{array}\right)}
\def\barraynb{\begin{array}}
\def\earraynb{\end{array}}

\def\IC{{{C}}} 
\def\IS{S} 
\def\SL{{\rm SL}}

\def\?{\quad{\gb{\fbox{\texttt{?}}\;}}\quad}

\def\v0{\mathbf{0}}

\usepackage{color}

\newcommand{\gb}{\colorbox{green}}

\def\beq{\begin{equation}}
\def\eeq{\end{equation}}
\def\bea{\begin{eqnarray}}
\def\eea{\end{eqnarray}}

\def\6{\partial}
\def\7{\tilde}
\def\8{\widehat}
 \def\bx{{\bf x}}

\def\G11{\Gamma_{11} }

\newcommand{\half }{\frac{1}{2}}

\let\ssection=\section
\renewcommand{\section}{\setcounter{equation}{0}\ssection}

\begin{document} 

\preprint{arXiv:1402.5894v3 [gr-qc]
}

\title{Conformal Carroll groups and BMS symmetry
\\[6pt]
}

\author{
C. Duval$^{1}$\footnote{Aix-Marseille Universit\'e, CNRS, CPT, UMR 7332, 13288 Marseille, France.
Universit\'e de Toulon, CNRS, CPT, UMR 7332, 83957 La Garde, France.
mailto:duval@cpt.univ-mrs.fr},
G. W. Gibbons$^{2,3,4}$\footnote{
mailto:G.W.Gibbons@damtp.cam.ac.uk},
P. A. Horvathy$^{3}$\footnote{mailto:horvathy@lmpt.univ-tours.fr},
}

\affiliation{
$^1$Centre de Physique Th\'eorique,
Marseille, France
\\
$^2$D.A.M.T.P., Cambridge University, U.K.
\\
$^3$Laboratoire de Math\'ematiques et de Physique
Th\'eorique,
Universit\'e de Tours,
France
\\
$^4$LE STUDIUM, Loire Valley Institute for Advanced Studies, Tours and Orleans France
}

\date{\today}


\begin{abstract} 
The Bondi-Metzner-Sachs (BMS) group is shown to be the conformal extension of L\'evy-Leblond's ``Carroll" group. Further extension to the Newman-Unti (NU)
 group is also discussed in the  Carroll  framework.
\end{abstract}

\maketitle


\section{Introduction}\label{Intro}

There has recently been a resurgence of interest in both  the Bondi-Metzner-Sachs (BMS) group \cite{BMS,Barnich,Bagchi,Strominger}) and, independently,  in L\'evy-Leblond's ``Carroll'' group \cite{Leblond, DGH, DGHZ}.


The BMS group arose as the asymptotic symmetry group
of a four dimensional  asymptotically flat spacetime representing an isolated
time-dependent system emitting gravitational radiation \cite{BMS}. It had been
anticipated, as is the case for asymptotically flat
time independent isolated systems, that the asymptotic symmetry group
would be the Poincar\'e  group and it came as a surprise       
that in the presence of gravitational radiation it is impossible
to isolate a unique asymptotic Poincar\'e group, but rather
the asymptotic symmetries constitute an infinite-dimensional
group which contains many copies of the  Poincar\'e group, none of which being invariant.

The BMS group may be thought of
as acting on  future (or past)  null infinity  ${\cal I}^{\pm}$.  
The latter are null hypersurfaces 
contained in the conformal boundary of an asymptotically flat spacetime.
Topologically  $ {\cal I}^{\pm} \equiv S^2 \times {\mathbb R} $, where
the $S^2$ factor corresponds to the 2-sphere of asymptotic directions
and  the ${\mathbb R}$ factor to retarded (advanced) time. Thus 
 local charts $\theta,\phi,u$ or  $\theta,\phi,v$ 
may be introduced, where $\theta,\phi$ parametrize the null generators 
and $u,v$ are affine parameters  along the null generators. From now on
we shall consider only ${\cal I}^{+}$ since the story for ${\cal I}^{-}$
is identical.

\goodbreak

Let us recall that ${\cal I}^{+}$ admits a degenerate conformal structure for which 
we may take a representative  metric of the form
$
ds^2 = 0 \times du^2 + d \theta ^2 + \sin^2 \theta d \phi^2 \,,
$    
where the non-vanishing summand is the standard round metric
on the unit 2-sphere which we wish to think of as the Riemann sphere ${\mathbb C} \cup \{\infty\}$.
To this end we introduce 
stereographic coordinates $ \zeta = e^{i\phi} \cot ({\theta}/{2} ) $ 
in terms of which 
$ ds^2 =
4\bigl( 1+ \zeta \bar \zeta \bigr )^{-2}d \zeta d \bar \zeta.
$  
The identity component of the  Lorentz  group is isomorphic to 
$\PSL(2,\bbC)=\SL(2,\bbC)/{\bbZ}_2$
and acts as conformal transformations of the Riemann sphere.  
Specifically, if $a,b,c,d \in {\mathbb C}$ are such that $ad-bc =1$, then for
\begin{equation}
\zeta' =\varphi({\zeta})
= 
\displaystyle\frac{a\zeta+b}{c \zeta+d},
\label{I.3}   
\end{equation}
we have 
\beq
\displaystyle\frac{d\zeta'd\bar\zeta'}{
\bigl(1+ \tilde\zeta\bar{\tilde\zeta} \bigr)^2}=
\Omega^2 (\zeta,\bar\zeta) 
\displaystyle\frac{d\zeta d \bar\zeta}{
 \bigl(1+\zeta \bar \zeta \bar)^2 },\,
\label{I.5}
\eeq
where
\beq
\Omega(\zeta,\bar\zeta)=\displaystyle\frac{1+\zeta\bar\zeta}{|a\zeta +b|^2 + |c\zeta +d|^2}.
\label{I.5bis}
\eeq
The infinite-dimensional abelian  group 
$\cT$ of \emph{super-translations}  
 acts on the 2-sphere sections (known as ``cuts'') of the product  as
$u\to{}u'= u + \alpha(\zeta,\bar\zeta)$ preserving the conformal structure, where  $\alpha(\zeta,\bar\zeta)$  is a smooth real valued function 
on the Riemann sphere which  transforms as a scalar of weight $1$ 
under conformal transformations of $S^2$ \cite{Geroch,McCarthy}.  The standard BMS group is thus 
the semi-direct product of $\mathrm{PSL}(2,\bbC)$ with 
$\cT\equiv{}C^\infty(S^2,\bbR)$ and acts on ${\cal I}^{+}$ as 
$(\zeta,u)\mapsto(\zeta',u')$ with $\zeta'$ as in Eq. (\ref{I.3}), and
\begin{equation}
u'= \Omega(\zeta,\bar{\zeta})\big[u + \alpha(\zeta,\bar\zeta)\big].
\label{I.6} 
\end{equation}
If  $\alpha$ is expanded in spherical harmonics and 
only the $l=0$ and $l=1$ terms retained, one obtains a closed subgroup
isomorphic with the Poincar\'e group. Now, the sum
of the  $l=0$ and $l=1$ representations of $\SO(3)$ transform
as the $D^{(\half\half )}$ representation of $\PSL(2,\bbC)$, i.e., as
the defining representation of the Lorentz group and as a scalar
of weight 1 under conformal transformations. 
However the Poincar\'e subgroup so defined is not an invariant
subgroup \cite{PenroseRindler}.  
In that case the generalized BMS group (and also its Lie algebra of vector fields) will be much larger than the Poincar\'e subgroup  defined above. 
 
Rather weaker versions of the BMS group  may be defined. For example,
the \emph{Newman-Unti (NU) group} is defined  by choosing
$a,b,c,d \in\bbC$ such that $ad-bc=1$, and set
\bea 
\zeta' = \frac{a\zeta+b}{c \zeta +d}, 
\quad 
u' =  f(\zeta,\bar\zeta,u),
\label{I.7}   
\eea
 where $f$ must decrease with $u$ at fixed $\zeta$ \cite{PenroseRindler}.
The NU group preserves the conformal geometry, but not what
is called the \emph{strong conformal geometry} \cite{PenroseRindler}. 
There is also an intermediate
version, for which
\beq
f(\zeta,\bar\zeta,u) 
=\beta(\zeta,\bar\zeta)\big[u+\alpha(\zeta,\bar\zeta)\big].
\label{I.9} 
\eeq 
Taking $\beta(\zeta,\bar \zeta)=\Omega(\zeta,\bar \zeta)$ gives the general element of the BMS group.

 The obvious generalization   is to replace $S^2$ by~$S^d$ and $\rO(3,1)$ by $\rO(d+1,1)$ \cite{Awada}, but since the asymptotics of solutions of the Einstein's equations  
in greater than four spacetime dimensions differ somewhat from the 
four-dimensional case \cite{Hollands}, its physical relevance remains unclear.
More generally, one may clearly consider the group obtained
by replacing the Riemann sphere by any closed 
 Riemannian  manifold
$(\Sigma,\hrg)$.
In general this will have no proper conformal isometries 
and so  one gets the semi-direct product of the isometry
group of $(\Sigma,\hrg)$ with $C^\infty(\Sigma,\bbR)$.
If $\Sigma$ is non-compact, for example the Euclidean
plane ${\mathbb E}^2\equiv\bbC$, then if one is unconcerned about global
issues, in a formal sense any  holomorphic map  
is conformal.
Similarly for the cylinder 
${\mathbb E}^2\setminus\!\{0\}\equiv\bbC\setminus\!\{0\}\equiv{}S^1\times\bbR$. 
In that case the generalized BMS group and its Lie algebra will be much larger
than that defined above. It is this type of infinite-dimensional extension of the standard BMS group which has figured in much of the current literature~\cite{Barnich,Bagchi}.

Our purpose being to relate all these groups to some particular overgroups of the \emph{Carroll group}, let us recall that the latter, 
originally introduced as an unusual contraction of the Poincar\'e group~\cite{Leblond}, is highlighted by that a ``Carrollian boost" with parameter $\bb\in\bbR^3$ transforms ``Carrollian time''~$u$ alone, according to
\beq
\bx'=\bx,
\qquad
u'=u-\bb\cdot\bx
\label{Carrboost}
\eeq
where $\bx\in\bbR^3$,
instead of the familiar Galilean action
$\bx'=\bx+\bb t$, $t'=t$ on non-relativistic spacetime.

The aim of the present Letter is thus to show that the BMS and NU groups can be understood as \emph{conformal symmetries},   
namely as \emph{conformal Carroll groups} introduced in this paper, associated with Carroll manifolds \cite{DGH,DGHZ}. Further extension to the \emph{Newman-Unti group} \cite{PenroseRindler} is also  discussed.

\section{Carroll group and manifold}

 In \cite{DGHZ}  a  definition independent of relativity and group contraction has  been put forward.  It is based on defining  a \emph{Carroll mani\-fold} analogous to a Newton-Cartan  manifold \cite{DHGalConf,DGHZ} as
a $(d+1)$-dimensional manifold $C$ (for Carroll), endowed with a twice-symmetric covariant, positive tensor field, $\rg$, whose kernel is generated by a nowhere vanishing vector field $\xi$.
The stronger definition proposed in Ref. \cite{DGHZ} requires a Carroll manifold to carry, in addition, a sym\-metric affine con\-nection, compatible with both $\rg$ and~$\xi$. The degeneracy of the ``metric'' $\rg$ implies that the con\-nection 
is not uniquely defined 
or may let alone exist, see 
Section \ref{Lightcone} 
below. In this Letter we stick essentially to our new and less restrictive definition given here.

(i) The standard Carroll structure is given  by 
$C^{d+1}=\bbR^d\times\bbR$, with $\rg=\delta_{AB}\,dx^A{}dx^B$, and
$\xi={\partial}/{\partial u}$, 
where $A, B = 1,\dots,d$ are spatial indices, and $u=x^{d+1}$ is the ``Carrollian ``time''-coordinate with dimension action/mass.


(ii) More general Carroll manifolds can be constructed  out of hypersurfaces~$\Sigma$ with Rieman\-nian metric $\hrg$,
 namely as
$C=\Sigma\times\bbR$ and $\rg=\hrg_{AB}(x)\,dx^A{}dx^B$, and
$\xi={\partial}/{\partial u}$.
In  both  cases, $C$ can also be endowed with a compatible connection, e.g., 
$\Gamma^k_{ij}=0$ for all $i,j,k=1,\ldots,d+1$ in  case~(i). The non-vanishing components of the Carroll connection of case (ii) being  
$\Gamma_{AB}^C=\hGamma_{AB}^C$ (the 
Levi-Civita connection of $\hrg$), the $\Gamma_{AB}^u$ remaining arbitrary.
Yet another important example will be studied 
in Section \ref{Lightcone} 
below.

\section{Conformal Carroll transformations}

Inspired by the definition of relativistic, and also non-relativistic \cite{DHGalConf}, conformal transformations, we now introduce for, a given Carroll manifold, the \textit{conformal Carroll group  of level $N$}, $\CCarr_N(C,\rg,\xi)$,  
of those transformations which preserve the tensor field 
$
\rg\otimes\xi^{\otimes{}N}
$
canonically associated with our Carroll structure, i.e., of all transformations, $a$, which satisfy 
\beq
a^*\rg=\Omega^2\rg
\quad \&\quad
a^*\xi=\Omega^{-2/{N}}\xi
\label{ConCarrN}
\eeq
 for some positive function $\Omega$  on $C$ and positive integer $N$. The Lie algebra  of infinitesimal \emph{conformal Carroll transformations}, $\confcarr_N(C,\rg,\xi)$, is spanned, accordingly, by vector fields $X$ such that
\begin{equation}
L_X\rg=\lambda\,\rg
\quad
\&
\quad
L_X\xi=-\frac{\lambda}{N}\,\xi
\label{confcarr}
\end{equation}
for some  function $\lambda$ on $\IC$.
 In the flat case (i) our formulae yield
 \cite{DGHProg}
\begin{eqnarray}
\nonumber
X
&=&
\Big(\omega^A_B\,x^B+\gamma^A+(\chi-2\kappa_Bx^B){}x^A\\
\nonumber
&&
+\kappa^Ax_Bx^B\Big)\frac{\partial}{\partial{}x^A}+\\
&&
\Big(\smallover{2}/{N}(\chi-2\kappa_Bx^B)u+T(x)
\Big)\frac{\partial}{\partial{}u}\qquad\;\;\;
\label{confcarrN}
\end{eqnarray}
where $\bomega\in\soo(d),\bgamma,\bkappa\in\bbR^d$, and $\chi\in\bbR$, with $T\in{}C^\infty(\bbR^d,\bbR)$. In view of (\ref{ConCarrN}), the supertranslation $T$ in (\ref{confcarrN}) has conformal weight $-2/N$, and should therefore be regarded as a density with weight $\nu=-2/(Nd)$.
Hence, $\confcarr_N(d+1)$ is the semi-direct product of the conformal Lie algebra $\soo(d+1,1)$ with $\nu$-densities
on $\bbR^d$ \footnote{The canonical $\nu$-densities of a Rieman\-nian manifold $(\Sigma,\hrg)$ are locally of the form $f\det{(\rg_{AB})}^{\nu/2}$ with~$f$ a smooth function on $\Sigma$.}. 
Due to the degeneracy of the Carroll ``metric'', $\rg$, our  conformal Carroll Lie algebras are \emph{infinite-dimensional}, 
owing to super-translations represented by $T$. In view of (\ref{confcarrN}), the quantity 
$z={2}/{N}$ is the associated \emph{dynamical exponent}.
The  value $N=2$ is particularly interesting: space and ``time'' are then equally dilated so that the dynamical exponent is $z=1$ \footnote{ 
 The special value $z=1$ corresponds to group contraction from the relativistic conformal group, analogous to that in the Galilean case \cite{CGA}. Accordingly, the conformal Carroll invariant  
$\rg\otimes\xi\otimes\xi$
can be regarded as a limit as $c\downarrow0$ of the conformal invariant $c^2\, G\otimes{}G^{-1}$ of a Lorentz manifold $(M,G)$ for which the Carroll manifold is an ultra-relativistic asymptote.}.
Moreover, in $d=1$ space dimension, interchanging position and time, $x\leftrightarrow{}s$ and re\-naming~$s$ as $t$,
our conformal Carroll algebra becomes precisely the Conformal Galilei algebra CGA~\cite{CGA}, \cite{DHGalConf}. Note that the interchange also swaps Carrollian (\ref{Carrboost}) and ordinary Galilean boosts.

 Requiring $\Omega=1$, i.e., $\lambda=0$ in (\ref{confcarr}) would  yield  the ``isometry group'' of the Carrollian  structure ($\rg,\xi$). It is infinite-dimensional owing to the presence
 of super\-translations. 
 Requiring, in addition, the preservation of a Carroll connection would reduce this to a finite-dimensional group; for, e.g., the flat Carroll structure (i) we get the usual \textit{Carroll group} 
\cite{Leblond}, denoted by $\Carr(d+1)$ in \cite{DGHZ}.
 The Carroll Lie algebra, $\carr(d+1)$, is spanned by the vector fields
\begin{equation}
X=(\omega^A_B\,x^B+\gamma^A)\frac{\partial}{\partial{}x^A}+(\sigma-\beta_A\,x^A)\,\frac{\partial}{\partial{}u},\quad
\label{carralg}
\end{equation}
where $\omega\in\soo(d)$, $\bbeta,\bgamma\in\bbR^d$, and $\sigma\in\bbR$.

In case (ii), the general expression of a conformal Carroll vector field in $\confcarr_N(C,\rg,\xi)$ is 
\beq
 X=Y+
\Big(\frac{\lambda}{N}\,u+T(x)\Big)\frac{\partial}{\partial{}u},
\label{ccSigma}
\eeq
where $\hX=\hX^A(x)\partial/\partial{x^A}$
is a conformal vector field of $(\Sigma,\hrg)$, i.e., such that $L_{\hX}\hrg=\lambda\,\hrg$, hence with $\lambda=(2/d)\widehat\nabla_AY^A$, 
and $T$ is a real function on $\Sigma$. 
Integration of the vector field (\ref{ccSigma}) readily yields the group action $(x,u)\mapsto(x',u')$, 
where 
\beq
x'=\varphi(x), 
\;
u'=\Omega^{2/N}(x)\big[u+\alpha(x)\big],
\label{TSigma}\eeq
with $\varphi\in\Conf(\Sigma,\hrg)$, and $\alpha\in{}C^\infty(\Sigma,\bbR)$.
Putting $a=(\varphi,\alpha)$, we readily find the group law of the conformal Carroll group of level $N$; if $a''=a'a$, we end up with the group law
\beq
\varphi''=\varphi'\circ\varphi,
\quad
\alpha''=\Omega^{-2/N}\varphi^*\alpha'+\alpha.
\label{grouplaw}
\eeq
 Notice, again, that supertranslations are actually densities of conformal weight $-2/N$. 
The conformal Carroll transformations of $C=\Sigma\times\bbR$ belong therefore to the \emph{semi-direct product of conformal transformations of $(\Sigma,\hrg)$ with ``supertranslations'' of~$\Sigma$}, 
\beq
\label{confcarrMxR}
\CCarr_N(C,\rg,\xi)\equiv
\Conf(\Sigma,\hrg)\ltimes{}\cT,
\eeq
where $\cT$ is a shorthand for  supertranslations
(mathematically, $-2/(Nd)$-densities on $\Sigma$).

If, for example, $\Sigma=\IS^1$ and $\rg=d\theta^2$,
conformal Carroll transformations of level $N$ will be given by the semi-direct product of the conformal transformations of the circle, $\Diff(S^1)$, and super\-translations with weight $\nu=-2/N$.
They are generated by the vector fields 
\beq
X=\hX(\theta)\frac{\partial}{\partial\theta}+\Big(\frac{2}{N}\hX'(\theta)\,u+T(\theta)\Big)\frac{\partial}{\partial{u}}\,.\qquad
\eeq

Considering instead $\Sigma=\IS^2$ endowed with its round  metric 
 allows us to conclude that the conformal Carroll transformations of level $N=2$ are the semi-direct product of the conformal group of $S^2$ with super\-translations, 
\begin{equation}
\CCarr_2(S^2\times\bbR,\rg,\xi)\equiv\PSL(2,\bbC)\ltimes{}
\cT\qquad
\label{confcarrS2xR}
\end{equation}
with $\cT$ the $-1/2$ densities on the two-sphere.
This is, precisely, the 
\emph{Bondi-Metzner-Sachs group $\BMS(4)$}~\cite{BMS} whose group law is given by (\ref{grouplaw}). Its action on our Carroll manifold can be read off from 
 (\ref{TSigma}) with $N=2$, and yielding Eqs (\ref{I.3}) and (\ref{I.6}).

Then, for $\Sigma=S^d$, the group $\SL(2,\bbC)$ is simply replaced by (the neutral component of) $\rO(d+1,1)$, and  Carroll isometries  readily identified with the semi-direct product of the
orthogonal group $\rO(d+1)$ with  supertranslations ---  and is therefore still infinite-dimensional. We therefore claim that $\CCarr_2(C,\rg,\xi)\equiv\BMS(d+2)$. 

\section{The light-cone as a Carroll manifold and the BMS group}\label{Lightcone}

To present our third example of a Carroll manifold, we deal with Minkowski spacetime 
$\bbR^{d+1,1}$ endowed with the metric
  $G=\mathrm{diag}(1,\ldots,1,-1)$, and look at the punctured future light cone, $C={\cal I}^{+}$, of  future pointing null vectors,  
 described (non canonically) by those vectors $(\bz,t)\in\bbR^{d+1}\times\bbR$ such that $t\equiv|\bz|>0$. 
Let us then consider the symmetric tensor, $\rg$, inherited on  $C$ from the Minkowski metric $G$. Using coordinates, and since $t>0$ can be viewed as a (global) radial coordinate on $C$, let the unit vector $\bx=\bz/t$ denote the direction of $\bz$. Then 
$ 
\rg=t^2\vert{}d\bx\vert^2=t^2\,\hrg,
$ 
where $\hg$ is the usual round metric of $S^d$ (the directions of light rays).
Let us insist that, the above decomposition of null vectors being non-canonical, ``projecting'' $\rg$ to the ``celestial sphere'', $S^d$,  
only defines a \emph{conformal class}, $[g]$, of  metrics of the latter. 
The kernel of $\rg$ is seen to be spanned by the restriction, $\xi$, to $C$ of the Euler vector field of $\bbR^{d+1,1}$, namely
$\xi=t\,{\partial}_{t}={\partial}_{u}$, where $u=\log t$. 

The light-cone $C={\cal I}^{+}$ is, hence, an \textit{intrinsical\-ly} defined Carroll manifold. Can it be endowed with a compatible  connection~?
Crucially for us, the answer is \emph{no~!}  To see this, choose a coordinate system $(x^A,u)$ on~$C$, such that
$ g_{AB}=e^{2u}\hg_{AB}$,
$g_{Au}=0$,
$g_{uu}=0$,
$\xi^A=0$,
$\xi^u=1$. 
Then a symmetric affine connection $\nabla$ on $C$  should  satisfy both conditions
$(\nabla_u{g})_{AB}=(\nabla_Ag)_{uB}=0$,
which are readily seen to be contradictory. 

 Providentially enough, our definition for a conformal Carroll transformation does not involve the connection, though. Then a calculation analogous to the proof of Eq. (\ref{confcarrMxR}) in the case~(ii) shows that $L_X\rg=\lambda\,\rg$ requires $\partial_uX^A=0$ as well as
$L_{\hX}\,\hrg=(\lambda-2X^u)\,\hrg$. 
Using the second condition
$L_X\xi=-(\lambda/N)\xi$ in (\ref{confcarr}) allows us to deduce that the conformal Carroll group 
of the punctured future light cone is
(\ref{confcarrS2xR}), i.e., for $N=2$, the 
\emph{BMS group}. \footnote{As explained before, for $d=1$ and $N=2$ the conformal Carroll group is the same as CGA with $z=1$, as seen by interchanging  position and time, cf. \cite{Bagchi}.}

The condition  
$\lambda=0$ would
\emph{fix the supertranslations} as $T=X^u=-\half(L_{\hX}\,\hrg)/\,\hrg$, while leaving the space-part, $\hX$, conformal; the Carroll ``isometries'' of the light-cone span therefore the conformal group, $\rO(d+1,1),$ of the celestial sphere, $S^d$, with rigidly fixed ``compensating'' supertranslations.

\section{Newman-Unti groups of the light-cone}

Our formalism allows us to define the \textit{Newman-Unti group} of a Carroll manifold, and of the light-cone in particular.

The Newman-Unti (NU) group is spanned of those (local) diffeomorphisms~$a$ of~$C$ which preserve the sole degenerate ``metric'', $\rg$, up to a conformal factor, namely
$
a^*[\rg]=[\rg]. 
$
This entails that the direction of $\xi$ is automatically preserved (since $a_*\xi$ lies again in the kernel of $\rg$). Its Lie algebra consists, hence, of all vector fields~ $X$ on~$C$ such that
$L_X\rg=\lambda\,\rg$, the condition $L_X\xi=\mu\,\xi$ being automatically satisfied.

If $C$ is the light-cone $\cI^+$ of $\bbR^{d+1,1}$, we  find that 
\begin{equation} 
X=\hX+X^u(x,u)\frac{\partial}{\partial{}u},
\label{nu}
\end{equation}
with
$\hX=\hX^A(x)\partial/\partial{x^A}$ a conformal vector field of $S^d$
and
$
X^u\in{}C^\infty(C,\bbR),
$
is, this time, an arbitrary function of the~$x^A$ \emph{and} $u$. 
The Newman-Unti group of the light-cone $C$ 
is, therefore,
\begin{equation}
\NU\equiv\Conf(S^d)\ltimes{}C^\infty(C,\bbR),
\label{NU}
\end{equation} 
consistently with Eqs (\ref{I.7}) and (\ref{I.9}). 
 We notice that  Eq. (\ref{I.9})  corresponds in fact to the ``intermediate''  Lie subalgebra of the NU Lie algebra defined by
$L_X\rg=\lambda\,\rg$,
and 
$(L_\xi)^2X=0$. 
For the light-cone, it consists in those vector fields $X=\hX+X^u\,\partial/\partial{}u$ with, just as before, $\hX\in\conf(S^d)$, and $(\partial_u)^2X^u=0$, i.e., such that $X^u=S\big[u+T\big]$,
where $S$, and $T$ remain arbitrary functions on $S^d$, see  Eq. (\ref{I.9}).

Referring to Eqs (\ref{ccSigma}) and (\ref{nu}) giving the generators of the conformal Carroll Lie algebras previously studied, we can highlight, in the case $N=2$, the interesting array of nested Lie groups \footnote{The $X^u$ component of the generators (\ref{nu}) is a polynomial of degree $k-1$ in $u$.}
\beq
\NU_1\subset\BMS\subset\NU_2\subset\cdots\subset\NU.
\label{nested}
\eeq 
We finally notice that $\NU_1=\CCarr_\infty$.

\goodbreak

\section{Concluding remarks}

The basic result of this paper
is that the future null conformal boundary ${\cal I}^+$  of an asymptotically
flat spacetime emitting gravitational radiation is a Carroll manifold \cite{DGHZ}
and its asymptotic symmetries, i..e.,  elements of  
the Bondi-Metzner-Sachs group~\cite{BMS},
constitute  the associated  conformal Carroll group. Originally introduced
as the limit of the Poincar\'e group as the speed of light tends
to zero \cite{Leblond}, Carroll groups and Carroll manifolds 
have found applications in the study
of velocity-dominated spacetimes and physics on  branes  which approach the 
speed of light. They are the analogue of Newton-Cartan
manifolds, which arise when when  the speed of light tends
to infinity~\cite{DGHZ}.  As a null hypersurface  in the conformal
compactification of spacetime ${\cal I} ^+$ carries an induced metric, $\rg$,
which is degenerate with a one-dimensional kernel spanned by $\xi$ which is tangent to its null generators. It also carries a so-called strong conformal structure \cite{PenroseRindler}.
Depending upon how much of this structure one requires to 
be preserved one may obtain different symmetry groups. Dropping the requirement
that the strong conformal structure be preserved leads to a larger group, namely to the Newman-Unti group \cite{PenroseRindler}, which is now seen to fit into the general
theory of Carroll manifolds and their symmetries.  

As we recalled in the introduction,
there has been a considerable revival of interest recently in
the BMS group in connection with its application to  
conformal field theory \cite{Barnich,Bagchi,Strominger}. We hope 
that the clarification  brought about in the present  paper will further advance this study. 

\goodbreak  

\begin{acknowledgments}
We are 
indebted to G.~Barnich whose advice allowed us to correct an important misinterpretation.
C.D. acknowledges ancient and enlightening discussions with F.~Ziegler about the BMS group. 
GWG would like to thank  KITP, Santa Barbara for its hospitality
during its {\it Bits and Branes} program (2012),
which provided a stimulus for this work. He is grateful  also to the 
{\it Laboratoire de Math\'ematiques et de Physique Th\'eorique de l'Universit\'e de Tours}  for hospitality, and the  {\it R\'egion Centre} for a \emph{
``Le Studium''} research professor\-ship. 
\end{acknowledgments}



\end{document}